\documentclass[manuscript,natbib,screen,review=false]{acmart}

\acmSubmissionID{760}

\usepackage[inline]{enumitem}
\usepackage{acronym}
\usepackage{xkcdcolors}

\PassOptionsToPackage{dvipsnames}{xcolor}

\usepackage{dsfont}
\usepackage[noend]{algorithmic}
\usepackage{algorithm}
\usepackage{bbm}
\usepackage{beramono}
\usepackage{bm}
\usepackage{booktabs}
\usepackage[skip=0pt]{caption}
\usepackage{cleveref}
\usepackage[T1]{fontenc}
\usepackage{graphicx}
\usepackage{hyperref}
\usepackage{listings}
\usepackage{multirow}
\usepackage{subcaption}
\usepackage{tabularx}
\usepackage[dvipsnames]{xcolor}
\usepackage{numprint}
\usepackage[normalem]{ulem}
\usepackage{multirow}
\usepackage{float}
\usepackage{multicol, blindtext}
\usepackage{mleftright}

\usepackage{subcaption}
\usepackage{adjustbox}

\acrodef{CF}{collaborative filtering}
\acrodef{LTR}{learning to Rank}
\acrodef{NDCG}{normalized discounted cumulative gain}
\acrodef{DCG}{Discounted Cumulative Gain}
\acrodef{VAE}{variational autoencoder}
\acrodef{ELBO}{Evidence Lower Bound Objective}
\acrodef{IPS}{inverse propensity scoring}
\acrodef{BPR}{Bayesian personalized ranking}
\acrodef{MF}{matrix factorization}
\acrodef{CRS}{conversational recommender system}
\acrodef{RS}{recommender system}
\acrodef{MNAR}{missing  not at random}
\acrodef{PE}{preference elicitation}
\acrodef{MSE}{mean squared error}
\acrodef{BINE}{bipartite network embedding}
\acrodef{MAE}{mean average error}
\acrodef{MSE}{mean squared error}
\acrodef{NDCG}{normalized discounted cummulative gain}
\acrodef{SNIPS}{self normalizing importance sampling}

\theoremstyle{definition}

\newcommand{\headernodot}[1]{\vspace{1mm}\noindent\textbf{#1}}
\newcommand{\header}[1]{\headernodot{#1.}}
\PassOptionsToPackage{hyphens}{url}\usepackage{hyperref}

\allowdisplaybreaks

\setcopyright{acmlicensed}
\copyrightyear{2023}
\acmYear{2023}
\acmDOI{}
\acmConference[CONSEQUENCES]{CONSEQUENCES Workshop at RecSys '23}{September 18-22}{Singapore}
\acmBooktitle{CONSEQUENCES Workshop at RecSys '23, September 18-22, 2023, Singapore}

\ccsdesc[300]{Information systems~Recommender systems}

\author{Shashank Gupta}
\affiliation{%
  \institution{University of Amsterdam}
  \city{Amsterdam}
  \country{The Netherlands}
}
\email{s.gupta2@uva.nl}

\author{Harrie Oosterhuis}
\affiliation{%
  \institution{Radboud Universiteit}
  \city{Nijmegen}
  \country{The Netherlands}
}
\email{harrie.oosterhuis@ru.nl}

\author{Maarten de Rijke}
\affiliation{%
  \institution{University of Amsterdam}
  \city{Amsterdam}
  \country{The Netherlands}
}
\email{m.derijke@uva.nl}

\AtBeginDocument{%
  }
    
\begin{document}

\begin{abstract}
 \Acl{PE} explicitly asks users what kind of recommendations they would like to receive. 
    It is a popular technique for conversational recommender systems to deal with cold-starts. 
    Previous work has studied selection bias in implicit feedback, e.g., clicks, and in some forms of explicit feedback, i.e., ratings on items.
    Despite the fact that the extreme sparsity of \acl{PE} interactions make them severely more prone to selection bias than natural interactions, the effect of selection bias in \acl{PE} on the resulting recommendations has not been studied yet.
    To address this gap, we take a first look at the effects of selection bias in \acl{PE} and how they may be further investigated in the future.
    We find that a big hurdle is the current lack of any publicly available dataset that has \acl{PE} interactions.
    As a solution, we propose a simulation of a topic-based \acl{PE} process.
    The results from our simulation-based experiments indicate 
    \begin{enumerate*}[label=(\roman*)]
    \item that ignoring the effect of selection bias early in \acl{PE} can lead to an exacerbation of overrepresentation in subsequent item recommendations, and 
    \item that debiasing methods can alleviate this effect, which leads to significant improvements in subsequent item recommendation performance. 
    \end{enumerate*}
    Our aim is for the proposed simulator and initial results to provide a starting point and motivation for future research into this important but overlooked problem setting.
\end{abstract}

\title[A First Look at Selection Bias in Preference Elicitation for Recommendation]{A First Look at Selection Bias in Preference Elicitation for Recommendation}

\maketitle

\section{Introduction}
\label{sec:intro}
Traditional recommender systems provide a single-shot human-system interface that is static in nature. 
They often rely on the user's past interactions to infer their preferences and generate a recommendation based on that. 
Traditional \ac{CF}-based methods fall into this category~\citep{ilievski2013personalized,jannach2018recommending,he2017neural}. 
However, these methods have trouble handling settings where user preferences are dynamic -- in practice, preferences often drift over time due to external covariates~\cite{jannach2018recommending} -- or single-shot recommendation settings where user intent has to be inferred from contextual information, instead of past interactions~\cite{mehrotra2019jointly}.
Additionally, these methods struggle to generate good recommendations for cold-start users and items. 
These issues, coupled with the sparse nature of user-item interaction data, make learning a good recommendation model difficult. 
A solution to these issues could be asking for a user's preferences directly at a coarser granularity in a \acfi{PE} stage.
Users are generally very willing to indicate or clarify their preferences, when prompted~\cite{priyogi2019preference}.

\begin{figure}[t]
	\centering
 \hfill
\includegraphics[scale=0.5,trim={5mm -12.5mm 0 0},clip]{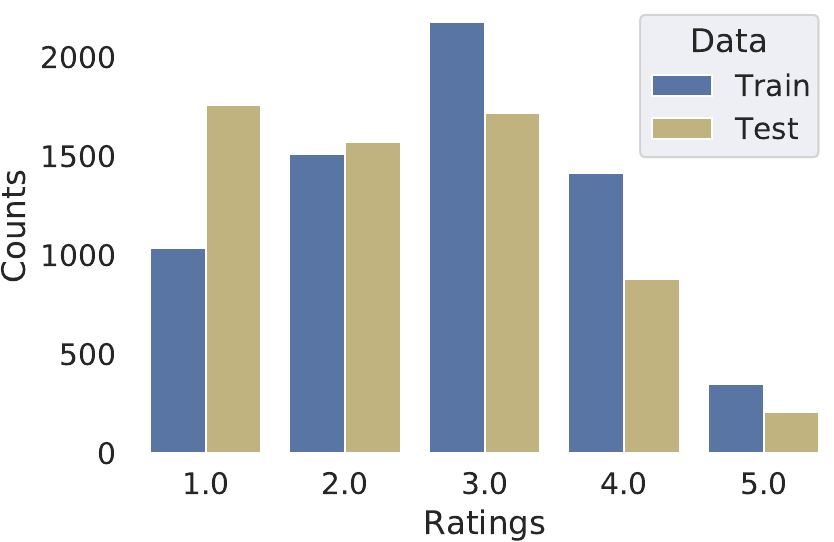}%
\hfill
\includegraphics[scale=0.495,trim={5mm 5mm 0 0},clip]{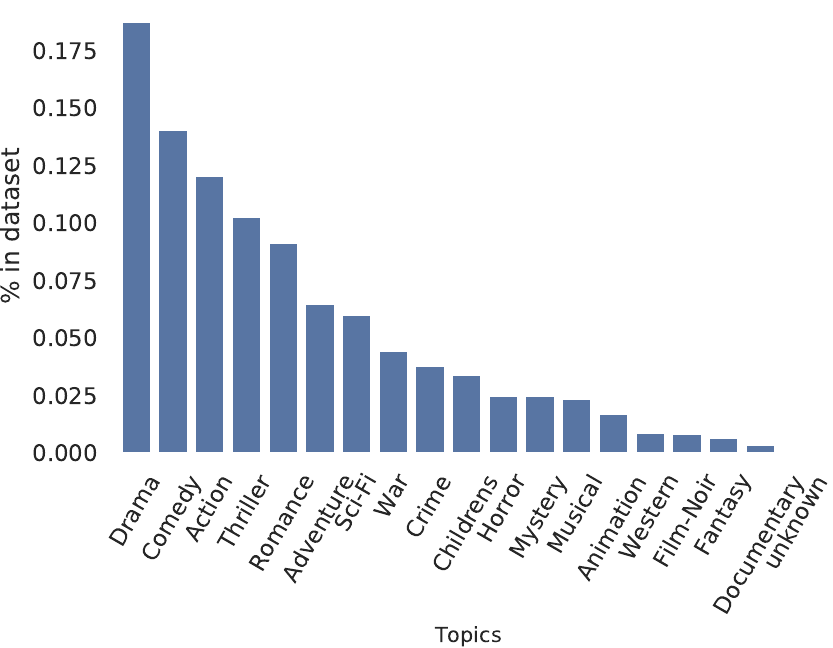}
\hfill
    \vspace{0.1\baselineskip}
    \caption{Rating distribution over item topics on the Coat Music dataset (Left), and Genre popularity in the MovieLens dataset (Right).
    }
	 \vspace{-3mm}
	\label{fig:rating-topic-plot} %
\end{figure}

\ac{PE} can be used in a variety of settings, including so-called ques\-tion-based \acp{CRS}~\citep{zhang2020task,christakopoulou2018q,lei2020estimation}, which consist of the following main components: 
\begin{enumerate*}[label=(\roman*)]
\item \acfi{PE}, where the user's preferences on items or item topics are collected or elicited, and, subsequently,
\item \emph{item recommendation}, where the system generates recommendations for users, conditioned on their response during the \ac{PE} stage.
\end{enumerate*}
The interactive aspect of \acp{CRS} can help in dealing with dynamic user preferences and the lack of intent information. 
It can also help with the cold-start problem, by collecting user's preferences on a group of items, instead of on an item directly~\citep{chang2015using}.

Recommender systems are commonly optimized based on logged user interactions.
However, such interactions provide a biased view of the actual user preferences~\citep{marlin2009collaborative,marlin2012collaborative,saito2020unbiased,yang2018unbiased}.
In particular, ratings are generally not evenly spread over all items but are heavily affected by popularity bias, resulting in a small number of items receiving most ratings.
Figure~\ref{fig:rating-topic-plot} (left) demonstrates this effect on the rating distribution of item topics in Coat, a popular recommendation dataset with an unbiased test set~\citep{schnabel2016recommendations}.
Popularity bias can be seen as a specific form of selection bias, due to which only part of the user preferences are observed in ratings~\citep{marlin2012collaborative}.
Importantly, selection bias on the item level propagates to the topic level; for example, Figure~\ref{fig:rating-topic-plot} demonstrates the popularity distribution over movie genres in the MovieLens dataset.
Similar to how selection bias in item ratings results in a biased view over topic preferences, it seems likely that selection bias in a \ac{PE} stage could negatively affect the subsequent recommendation stage.
While selection bias in user interaction data is widely studied~\citep{marlin2009collaborative,marlin2012collaborative,saito2020unbiased,yang2018unbiased,schnabel2016recommendations,marlin2012collaborative}, to the best of our knowledge, previous work has not considered the effects of selection bias in \ac{PE}.
To address this gap, this work takes a first look at the problem of selection bias in \ac{PE} for recommendation.
 We focus on elicitation on the topic-level followed by subsequent item recommendation.
 Because there is currently no publicly available recommendation dataset that represents \ac{PE}, we introduce a method for simulating a \ac{PE} stage from static recommendation datasets.
Our experimental results in the simulator reveal that selection bias in the \ac{PE} stage does, indeed, have negative effects on subsequent item recommendation.
We find that existing debiasing methods can be adapted to reduce these effects, leading to significantly better recommendations.

\vspace{-1mm}
\section{Correcting for Selection Bias in Preference Elicitation}
In this section, we discuss how common debiasing methods for item recommendation can be applied to topic-level \ac{PE}~\cite{schnabel2016recommendations}.
Let $U$ be the set of all users, $I$ the set of all items, and $T$ the set of all item-topics (referred to as topics hereafter) in the dataset, and $Y\in \{0,1\}^{|U| \cdot |T| }$ the user-topic \emph{complete} rating matrix; $Y_{u,t}$ is the true rating for the pair $(u,t)$.
$T\in \{0,1\}^{|I| \cdot |T| }$ is the indicator matrix where $T_{i,t} =1$ if item $i$ belongs to the topic $t$.
$R\in \{0,1\}^{|U| \cdot |I| }$ is the rating matrix, with entry $R_{u,i}$ indicating user $u$'s rating for item $i$. 
In reality, not all entries in the $Y$ matrix are observed; let $O\in \{0,1\}^{|U| \cdot |T| }$ be the observation matrix, with $O_{u,t}$ indicating whether the rating $Y_{u,t}$ is observed or not. 
The entries in the $Y_{u,t}$ matrix are affected by selection bias.
$O$ controls the selection bias, where certain ratings are overrepresented or underrepresented in the dataset; we use $\rho_{u,t} = P(O_{u,t} = 1)$ to denote the probability of observing a rating $Y_{u,t}$ in the dataset.

\header{Ideal rating estimator}
An ideal rating prediction loss can be defined as follows: 
\begin{equation}
    \mathcal{L}_\text{ideal} = \frac{1}{|U||T|} \sum_{u,t} L(\hat{y}_{u,t}, y_{u,t}).
    \label{eq:ideal_loss}
\end{equation}
The loss function $L(\hat{y}_{u,t}, y_{u,t})$ used for rating prediction could be \ac{MSE}.

\header{Naive rating estimator}
One could naively ignore selection bias in the observed rating data and estimate the prediction loss by simple averaging, resulting in the naive training loss estimator:
\begin{equation}
    \mathcal{L}_\text{naive} = \frac{1}{|\{u,t : O_{u,t}=1\}|} \sum_{u,t : O_{u,t}=1} L(\hat{y}_{u,t}, y_{u,t}),
    \label{eq:naive_loss}
\end{equation}
where $|\{u,t : O_{u,t}=1\}|$ is the number of observed ratings in the dataset. 
It is clearly a biased estimator of the ideal-loss (Eq.~\ref{eq:ideal_loss})~\cite{schnabel2016recommendations}.

\header{Unbiased preference elicitation}
To debias the loss function in Eq.~\ref{eq:naive_loss}, we apply \ac{IPS}~\citep{joachims2017unbiased,schnabel2016recommendations,saito2020unbiased}, where the propensity value $\rho_{u,t} = p(O_{u,t} =1)$ is used as a weight in the loss function.
The modified loss function is defined as follows:
\begin{equation}
    \mathcal{L}_\text{ips} = \frac{1}{|U||T|} \sum_{u,t : O_{u,t}=1} \frac{L(\hat{y}_{u,t}, y_{u,t})}{\rho_{u,t}}.
    \label{eq:ips_loss}
\end{equation}
The modified $\mathcal{L}_\text{ips}$ is an unbiased estimate of the ideal-loss defined in Eq.~\ref{eq:ideal_loss}
\cite{schnabel2016recommendations,saito2020unbiased}, i.e.,
$
    \mathbb{E}_{O} [\mathcal{L}_\text{ips}] = \mathcal{L}_\text{ideal}.
$

\vspace{-1mm}
\section{Experiments}

Below, we discuss the semi-synthetic experimental setup, fully-synthetic setup, followed by empirical results. For details on simulating preference elicitation data, and the synthetic topic generation, we defer to Appendix~\ref{exp-setup}.

\header{Yahoo! R3 dataset} This dataset is collected as part of a music-recommendation service; it includes rating information from 15,400 users on 1,000 items, which are self-selected by users, i.e., these are MNAR ratings~\citep{yahoo-r3}. A separate test-set comprises of ratings from a uniformly-random policy, ensuring the ratings are free from selection bias. 
Topic information is not present in the dataset, hence we use the synthetic topic generation method discussed in Appendix~\ref{sec:synt-topic}.
We use 20\% of the unbiased test data to generate the bipartite user-item graph and generate item embeddings, followed by synthetic topic generation, and finally the unbiased \ac{PE} data (Appendix~\ref{sec:synt-topic}). 
For clustering, we experiment with different numbers of clusters to evaluate the robustness of the method under different setups. 

\noindent\textbf{Fully-synthetic dataset.} Along with simulating conversations from user-item interactions, we also experiment with a fully-synthetic dataset setting,
where we simulate user-topic interactions directly. Following \citep{huang2020keeping}, the following two stage process is applied:
\begin{enumerate*}[label=(\roman*)]
\item Given $N$ users and $T$  topics, 
their corresponding latent-factors for users ($\mathbf{P}\in \mathbf{R}^{N * d} $) and topics ($\mathbf{Q}\in \mathbf{R}^{T * d} $) are generated via Gaussian distribution
$\mathcal{N}(0,\,1)$. The rating scores are generated via a dot-produce of user and topic latent factors. And
\item the MNAR logged data is generated via the following mechanism:
\end{enumerate*}
\begin{align}
    P(o_{u,t}
    \mid y_{u,t}) = \alpha P(o_{u,t}
    \mid y_{u,t}, \text{pos-bias}) + (1 - \alpha)P(o_{u,i}\mid \text{uniform}) \label{synt-eq}
\end{align} 
The simulator is available at: \url{https://github.com/shashankg7/Bias-Preference-Elicitation}.

\begin{table}[t]
    \begin{minipage}{.48\linewidth}
    \centering
    \caption{Performance of the debiasing method on the unbiased rating prediction task on the Yahoo! R3 dataset. Significant improvements over the baseline (MF) are marked with $^\dag$ ($p<0.01$). Average values over 10 different runs are reported.} %
    \label{table:result-yahoo}
    \vspace{0.2\baselineskip}
    \setlength{\tabcolsep}{1mm}
        \begin{tabular}{l l cc cc cc }
        \toprule
        \bf Exp. setting &  \bf Method & \multicolumn{1}{c}{\bf MAE$\downarrow$}    & \multicolumn{1}{c}{\bf MSE$\downarrow$} & \multicolumn{1}{c}{\bf NDCG@3$\uparrow$}   \\
        \midrule
        \multirow{3}{*}{$\# \text{clusters}=25$}  
        & MF & 1.3041 & 2.5634 & 0.7461 \\
        & ExpoMF & 1.3075 & 2.8213 & 0.7503 \\
        & MF-IPS & \textbf{0.8327}\rlap{$^\dagger$} & \textbf{1.0832}\rlap{$^\dagger$} & \textbf{0.7511}\rlap{$^\dagger$} \\
        \midrule
        \multirow{3}{*}{$\# \text{clusters}=50$}  
        & MF & 1.3094 & 2.5857 & 0.7476  \\
        & ExpoMF & 1.3050 & 2.8138 & 0.7511 \\
        & MF-IPS  & \textbf{0.8268}\rlap{$^\dagger$} & \textbf{1.0777}\rlap{$^\dagger$} & \textbf{0.7553}\rlap{$^\dagger$}  \\
        \midrule
        \multirow{3}{*}{$\# \text{clusters}=75$}  
        & MF & 1.3112 & 2.5887 & 0.7460  \\ 
        & ExpoMF & 0.8451 & 1.1530 & 0.7505 \\
        & MF-IPS & \textbf{0.8451}\rlap{$^\dagger$} & \textbf{1.1530}\rlap{$^\dagger$} & \textbf{0.7521}\rlap{$^\dagger$} \\
        \midrule
        \multirow{3}{*}{$\# \text{clusters}=100$}  
        & MF & 1.3057 & 2.5403 & 0.7460 \\
        & ExpoMF & 1.3109 & 2.8316 & 0.7499 \\
        & MF-IPS & \textbf{0.8464}\rlap{$^\dagger$} & \textbf{1.1553}\rlap{$^\dagger$} & \textbf{0.7518}\rlap{$^\dagger$}  \\
        \bottomrule
        \end{tabular}
    %
    \end{minipage}%
    \hfill
    \begin{minipage}{.48\linewidth}
    \centering
    \caption{Performance of the debiasing method on the unbiased rating prediction task on the fully-synthetic dataset.
    Significant improvements over the baseline (MF) are marked with $^\dag$ ($p<0.01$). Average values over 10 different runs are reported.
    }
    \label{table:result-synt}
    \vspace{0.2\baselineskip}
        \setlength{\tabcolsep}{1mm}
        \begin{tabular}{l l cc cc cc}
        \toprule
        \bf Exp.\ setting &  \bf Method & \multicolumn{1}{c}{\bf MAE$\downarrow$} & \multicolumn{1}{c}{\bf MSE$\downarrow$} & \multicolumn{1}{c}{\bf NDCG@3$\uparrow$} \\
        \midrule
        \multirow{3}{*}{$\alpha=0.25$}  
        & MF & 0.8449 & 1.0847  & \textbf{0.7611}  \\
        & ExpoMF & 1.6643 & 3.9344  & 0.6638  \\
        & MF-IPS & \textbf{0.7894}\rlap{$^\dagger$} & \textbf{0.9874}\rlap{$^\dagger$}  & 0.7511 \\
        \midrule
        \multirow{3}{*}{$\alpha=0.5$}  
        & MF & 0.8666 & 1.1461 & \textbf{0.7852}  \\
        & ExpoMF & 1.6506 & 3.9178 & 0.6838 \\
        & MF-IPS & \textbf{0.7670}\rlap{$^\dagger$} & \textbf{0.9185}\rlap{$^\dagger$}  & 0.7836  \\
        \midrule
        \multirow{3}{*}{$\alpha=0.75$}  
        & MF & 0.9012 & 1.2383 & 0.8053 \\
        & ExpoMF & 1.6469 & 3.9708 & 0.6984 \\
        & MF-IPS & \textbf{0.7330}\rlap{$^\dagger$} & \textbf{0.8322}\rlap{$^\dagger$} & \textbf{0.8230}\rlap{$^\dagger$} \\
        \midrule
        \multirow{3}{*}{$\alpha=1.0$}  
        & MF & 0.9622 & 1.3974 & 0.8179   \\
        & ExpoMF & 1.6473 & 4.0386 & 0.7121 \\
        & MF-IPS & \textbf{0.7254}\rlap{$^\dagger$} & \textbf{0.8078}\rlap{$^\dagger$}  & \textbf{0.8362}\rlap{$^\dagger$}   \\
        \bottomrule
        \end{tabular}
    \end{minipage} 
\end{table}

\section{Results}

We evaluate the effect of debiasing \ac{PE} on the unbiased test set. We use \acf{MAE} and \acf{MSE} as evaluation metrics \citep{schnabel2016recommendations} for measuring accuracy in rating prediction.
To evaluate the quality of rankings, we use \acs{NDCG}@3, following \citet{saito2020asymmetric}. We use ExpoMF \cite{liang2016modeling} as a baseline for debiasing, which uses a generative model to correct for the bias. 

Results for the semi-synthetic dataset are presented in Table~\ref{table:result-yahoo}.
Results are reported for different numbers of item clusters in the synthetic topic generation (see Section~\ref{sec:synt-topic}).
Different numbers of clusters represent a different \ac{PE} setting where the number of item topics varies. 
Metric values suggest that a naive method for learning rating prediction (using the objective in Eq.~\ref{eq:naive_loss}) results in sub-optimal performance across all settings of clusters. 
The results suggest that, even for a small-scale \ac{PE} system  (with 35 item-topics), a selection-bias exists, and using IPS for debiasing helps. 
For the fully-synthetic setup, results are presented in Table~\ref{table:result-synt}. 
Results are reported for different values of $\alpha$ (see Eq.~\ref{synt-eq}), which represent different levels of selection bias. 
A lower value of $\alpha$ represents a setting where the second term (with uniform observation probability) dominates, simulating a setting where data is sampled from a uniformly-random policy. 
Similarly, a higher $\alpha$ value represents a setting with higher positivity-bias. 
The value of $\alpha$ controls the degree of positivity bias in the simulated logged data. 
The results from a debiasing rating-prediction method (MF-IPS) are consistent with the results in the semi-synthetic setting for the rating prediction task, for the \ac{MAE} and \ac{MSE} metrics. 
However, for lower values of $\alpha$ (0.25, 0.5), the baseline \ac{MF} outperforms other methods in terms of NDCG.
We suspect this is caused by the uniform data generation part dominating the biased counterpart, hence there is less signal for learning user preferences. 
For higher $\alpha$ values, the results are consistently better for the IPS method. 
It is also interesting to note that even for the case where the uniformly-random policy dominates ($\alpha=0.25$), debiasing improves the performance in terms of \ac{MAE} and \ac{MSE}.

The results in this section show that a naive method for rating prediction in the \ac{PE} stage results in a sub-optimal system,
which we consistently observe across all experimental setups.

\vspace{-3mm}
\section{Conclusion}
We have explored the effect of selection bias in \ac{PE} for recommender systems.
We have shown that user-item interactions (ratings) in the preference elicitation stage suffer from the issue of selection bias, which is a common issue when dealing with ratings at the item-level~\citep{schnabel2016recommendations}. 
We have also explored how training a \ac{PE} system on biased data can lead to error propagation in downstream tasks. 
To the best of our knowledge, we are the first to explore and identify the issue of bias in the \ac{PE} stage. 
We have shown that, similar to the case of static item recommendations, selection bias exists in a \ac{PE} setting as well.

We have also investigated the application of existing debiasing methods used in item-based recommendation methods, and have shown that these methods can be successfully applied in our setting. 
Importantly, given a lack of unbiased test collections for evaluating bias in a \ac{PE}, we have proposed, and are sharing, a simulation method to generate an unbiased test collection for evaluating debiasing methods.
Finally, with the release of our simulator and experimental source code, in addition to our comparison of existing methods, we wish to provide a starting point and motivation for future research to further investigate the problem of bias in similar areas. As part of future work, we propose a joint debiasing method for the \ac{PE} stage and the corresponding downstream tasks. 

\bibliographystyle{ACM-Reference-Format}
\bibliography{references}


\begin{thebibliography}{23}


\ifx \showCODEN    \undefined \def \showCODEN     #1{\unskip}     \fi
\ifx \showDOI      \undefined \def \showDOI       #1{#1}\fi
\ifx \showISBNx    \undefined \def \showISBNx     #1{\unskip}     \fi
\ifx \showISBNxiii \undefined \def \showISBNxiii  #1{\unskip}     \fi
\ifx \showISSN     \undefined \def \showISSN      #1{\unskip}     \fi
\ifx \showLCCN     \undefined \def \showLCCN      #1{\unskip}     \fi
\ifx \shownote     \undefined \def \shownote      #1{#1}          \fi
\ifx \showarticletitle \undefined \def \showarticletitle #1{#1}   \fi
\ifx \showURL      \undefined \def \showURL       {\relax}        \fi
\providecommand\bibfield[2]{#2}
\providecommand\bibinfo[2]{#2}
\providecommand\natexlab[1]{#1}
\providecommand\showeprint[2][]{arXiv:#2}

\bibitem[Chang et~al\mbox{.}(2015)]%
        {chang2015using}
\bibfield{author}{\bibinfo{person}{Shuo Chang}, \bibinfo{person}{F~Maxwell
  Harper}, {and} \bibinfo{person}{Loren Terveen}.}
  \bibinfo{year}{2015}\natexlab{}.
\newblock \showarticletitle{Using Groups of Items for Preference Elicitation in
  Recommender Systems}. In \bibinfo{booktitle}{\emph{Proceedings of the 18th
  ACM Conference on Computer Supported Cooperative Work \& Social Computing}}.
  \bibinfo{pages}{1258--1269}.
\newblock


\bibitem[Christakopoulou et~al\mbox{.}(2018)]%
        {christakopoulou2018q}
\bibfield{author}{\bibinfo{person}{Konstantina Christakopoulou},
  \bibinfo{person}{Alex Beutel}, \bibinfo{person}{Rui Li},
  \bibinfo{person}{Sagar Jain}, {and} \bibinfo{person}{Ed~H. Chi}.}
  \bibinfo{year}{2018}\natexlab{}.
\newblock \showarticletitle{Q\&R: A Two-Stage Approach toward Interactive
  Recommendation}. In \bibinfo{booktitle}{\emph{Proceedings of the 24th ACM
  SIGKDD International Conference on Knowledge Discovery \& Data Mining}}.
  \bibinfo{pages}{139--148}.
\newblock


\bibitem[Gao et~al\mbox{.}(2018)]%
        {gao2018bine}
\bibfield{author}{\bibinfo{person}{Ming Gao}, \bibinfo{person}{Leihui Chen},
  \bibinfo{person}{Xiangnan He}, {and} \bibinfo{person}{Aoying Zhou}.}
  \bibinfo{year}{2018}\natexlab{}.
\newblock \showarticletitle{{BiNE}: Bipartite Network Embedding}. In
  \bibinfo{booktitle}{\emph{The 41st International ACM SIGIR Conference on
  Research \& Development in Information Retrieval}}.
  \bibinfo{pages}{715--724}.
\newblock


\bibitem[He et~al\mbox{.}(2017)]%
        {he2017neural}
\bibfield{author}{\bibinfo{person}{Xiangnan He}, \bibinfo{person}{Lizi Liao},
  \bibinfo{person}{Hanwang Zhang}, \bibinfo{person}{Liqiang Nie},
  \bibinfo{person}{Xia Hu}, {and} \bibinfo{person}{Tat-Seng Chua}.}
  \bibinfo{year}{2017}\natexlab{}.
\newblock \showarticletitle{Neural Collaborative Filtering}. In
  \bibinfo{booktitle}{\emph{Proceedings of the 26th International Conference on
  World Wide Web}}. \bibinfo{pages}{173--182}.
\newblock


\bibitem[Huang et~al\mbox{.}(2020)]%
        {huang2020keeping}
\bibfield{author}{\bibinfo{person}{Jin Huang}, \bibinfo{person}{Harrie
  Oosterhuis}, \bibinfo{person}{Maarten de Rijke}, {and} \bibinfo{person}{Herke
  van Hoof}.} \bibinfo{year}{2020}\natexlab{}.
\newblock \showarticletitle{Keeping Dataset Biases out of the Simulation: A
  Debiased Simulator for Reinforcement Learning based Recommender Systems}. In
  \bibinfo{booktitle}{\emph{Fourteenth ACM Conference on Recommender Systems}}.
  \bibinfo{pages}{190--199}.
\newblock


\bibitem[Ilievski and Roy(2013)]%
        {ilievski2013personalized}
\bibfield{author}{\bibinfo{person}{Ilija Ilievski} {and} \bibinfo{person}{Sujoy
  Roy}.} \bibinfo{year}{2013}\natexlab{}.
\newblock \showarticletitle{Personalized News Recommendation based on Implicit
  Feedback}. In \bibinfo{booktitle}{\emph{Proceedings of the 2013 International
  News Recommender Systems Workshop and Challenge}}. \bibinfo{pages}{10--15}.
\newblock


\bibitem[Jannach et~al\mbox{.}(2018)]%
        {jannach2018recommending}
\bibfield{author}{\bibinfo{person}{Dietmar Jannach}, \bibinfo{person}{Lukas
  Lerche}, {and} \bibinfo{person}{Markus Zanker}.}
  \bibinfo{year}{2018}\natexlab{}.
\newblock \showarticletitle{Recommending Based on Implicit Feedback}.
\newblock In \bibinfo{booktitle}{\emph{Social Information Access}}.
  \bibinfo{publisher}{Springer}, \bibinfo{pages}{510--569}.
\newblock


\bibitem[Joachims et~al\mbox{.}(2017)]%
        {joachims2017unbiased}
\bibfield{author}{\bibinfo{person}{Thorsten Joachims}, \bibinfo{person}{Adith
  Swaminathan}, {and} \bibinfo{person}{Tobias Schnabel}.}
  \bibinfo{year}{2017}\natexlab{}.
\newblock \showarticletitle{Unbiased Learning-to-Rank with Biased Feedback}. In
  \bibinfo{booktitle}{\emph{Proceedings of the Tenth ACM International
  Conference on Web Search and Data Mining}}. \bibinfo{pages}{781--789}.
\newblock


\bibitem[Kingma and Ba(2014)]%
        {kingma2014adam}
\bibfield{author}{\bibinfo{person}{Diederik~P. Kingma} {and}
  \bibinfo{person}{Jimmy Ba}.} \bibinfo{year}{2014}\natexlab{}.
\newblock \showarticletitle{Adam: A Method for Stochastic Optimization}. In
  \bibinfo{booktitle}{\emph{International Conference on Learning
  Representations, ICLR}}.
\newblock


\bibitem[Lei et~al\mbox{.}(2020)]%
        {lei2020estimation}
\bibfield{author}{\bibinfo{person}{Wenqiang Lei}, \bibinfo{person}{Xiangnan
  He}, \bibinfo{person}{Yisong Miao}, \bibinfo{person}{Qingyun Wu},
  \bibinfo{person}{Richang Hong}, \bibinfo{person}{Min-Yen Kan}, {and}
  \bibinfo{person}{Tat-Seng Chua}.} \bibinfo{year}{2020}\natexlab{}.
\newblock \showarticletitle{Estimation-Action-Reflection: Towards Deep
  Interaction between Conversational and Recommender Systems}. In
  \bibinfo{booktitle}{\emph{Proceedings of the 13th International Conference on
  Web Search and Data Mining}}. \bibinfo{pages}{304--312}.
\newblock


\bibitem[Liang et~al\mbox{.}(2016)]%
        {liang2016modeling}
\bibfield{author}{\bibinfo{person}{Dawen Liang}, \bibinfo{person}{Laurent
  Charlin}, \bibinfo{person}{James McInerney}, {and} \bibinfo{person}{David~M
  Blei}.} \bibinfo{year}{2016}\natexlab{}.
\newblock \showarticletitle{Modeling User Exposure in Recommendation}. In
  \bibinfo{booktitle}{\emph{Proceedings of the 25th international conference on
  World Wide Web}}. \bibinfo{pages}{951--961}.
\newblock


\bibitem[Marlin and Zemel(2009)]%
        {marlin2009collaborative}
\bibfield{author}{\bibinfo{person}{Benjamin~M. Marlin} {and}
  \bibinfo{person}{Richard~S. Zemel}.} \bibinfo{year}{2009}\natexlab{}.
\newblock \showarticletitle{Collaborative Prediction and Ranking with
  Non-Random Missing Data}. In \bibinfo{booktitle}{\emph{Proceedings of the
  Third ACM Conference on Recommender Systems}}. \bibinfo{pages}{5--12}.
\newblock


\bibitem[Marlin et~al\mbox{.}(2007)]%
        {marlin2012collaborative}
\bibfield{author}{\bibinfo{person}{Benjamin~M Marlin},
  \bibinfo{person}{Richard~S Zemel}, \bibinfo{person}{Sam Roweis}, {and}
  \bibinfo{person}{Malcolm Slaney}.} \bibinfo{year}{2007}\natexlab{}.
\newblock \showarticletitle{Collaborative Filtering and the Missing at Random
  Assumption}. In \bibinfo{booktitle}{\emph{Proceedings of the Twenty-Third
  Conference on Uncertainty in Artificial Intelligence}}.
  \bibinfo{pages}{267--275}.
\newblock


\bibitem[Mehrotra et~al\mbox{.}(2019)]%
        {mehrotra2019jointly}
\bibfield{author}{\bibinfo{person}{Rishabh Mehrotra}, \bibinfo{person}{Mounia
  Lalmas}, \bibinfo{person}{Doug Kenney}, \bibinfo{person}{Thomas Lim-Meng},
  {and} \bibinfo{person}{Golli Hashemian}.} \bibinfo{year}{2019}\natexlab{}.
\newblock \showarticletitle{Jointly Leveraging Intent and Interaction Signals
  to Predict User Satisfaction with Slate Recommendations}. In
  \bibinfo{booktitle}{\emph{The World Wide Web Conference}}.
  \bibinfo{pages}{1256--1267}.
\newblock


\bibitem[Priyogi(2019)]%
        {priyogi2019preference}
\bibfield{author}{\bibinfo{person}{Bilih Priyogi}.}
  \bibinfo{year}{2019}\natexlab{}.
\newblock \showarticletitle{Preference Elicitation Strategy for Conversational
  Recommender System}. In \bibinfo{booktitle}{\emph{Proceedings of the Twelfth
  ACM International Conference on Web Search and Data Mining}}.
  \bibinfo{publisher}{ACM}, \bibinfo{pages}{824--825}.
\newblock


\bibitem[Reynolds(2009)]%
        {reynolds2009gaussian}
\bibfield{author}{\bibinfo{person}{Douglas~A. Reynolds}.}
  \bibinfo{year}{2009}\natexlab{}.
\newblock \showarticletitle{Gaussian Mixture Models.}
\newblock \bibinfo{journal}{\emph{Encyclopedia of Biometrics}}
  \bibinfo{volume}{741}, \bibinfo{number}{659-663} (\bibinfo{year}{2009}).
\newblock


\bibitem[Saito(2020)]%
        {saito2020asymmetric}
\bibfield{author}{\bibinfo{person}{Yuta Saito}.}
  \bibinfo{year}{2020}\natexlab{}.
\newblock \showarticletitle{Asymmetric Tri-training for Debiasing
  Missing-not-at-random Explicit Feedback}. In
  \bibinfo{booktitle}{\emph{Proceedings of the 43rd International ACM SIGIR
  Conference on Research and Development in Information Retrieval}}.
  \bibinfo{pages}{309--318}.
\newblock


\bibitem[Saito et~al\mbox{.}(2020)]%
        {saito2020unbiased}
\bibfield{author}{\bibinfo{person}{Yuta Saito}, \bibinfo{person}{Suguru
  Yaginuma}, \bibinfo{person}{Yuta Nishino}, \bibinfo{person}{Hayato Sakata},
  {and} \bibinfo{person}{Kazuhide Nakata}.} \bibinfo{year}{2020}\natexlab{}.
\newblock \showarticletitle{Unbiased Recommender Learning from
  Missing-Not-At-Random Implicit Feedback}. In
  \bibinfo{booktitle}{\emph{Proceedings of the 13th International Conference on
  Web Search and Data Mining}}. \bibinfo{pages}{501--509}.
\newblock


\bibitem[Schnabel et~al\mbox{.}(2016)]%
        {schnabel2016recommendations}
\bibfield{author}{\bibinfo{person}{Tobias Schnabel}, \bibinfo{person}{Adith
  Swaminathan}, \bibinfo{person}{Ashudeep Singh}, \bibinfo{person}{Navin
  Chandak}, {and} \bibinfo{person}{Thorsten Joachims}.}
  \bibinfo{year}{2016}\natexlab{}.
\newblock \showarticletitle{Recommendations as Treatments: Debiasing Learning
  and Evaluation}. In \bibinfo{booktitle}{\emph{International Conference on
  Machine Learning}}. PMLR, \bibinfo{pages}{1670--1679}.
\newblock


\bibitem[Swaminathan and Joachims(2015)]%
        {swaminathan2015self}
\bibfield{author}{\bibinfo{person}{Adith Swaminathan} {and}
  \bibinfo{person}{Thorsten Joachims}.} \bibinfo{year}{2015}\natexlab{}.
\newblock \showarticletitle{The Self-normalized Estimator for Counterfactual
  Learning}. In \bibinfo{booktitle}{\emph{Proceedings of the 28th International
  Conference on Neural Information Processing Systems-Volume 2}}.
  \bibinfo{pages}{3231--3239}.
\newblock


\bibitem[{Yahoo! R3}(2022)]%
        {yahoo-r3}
\bibfield{author}{\bibinfo{person}{{Yahoo! R3}}.}
  \bibinfo{year}{2022}\natexlab{}.
\newblock \bibinfo{title}{R3 - Yahoo! Music Ratings for User Selected and
  Randomly Selected Songs, version 1.0}.
\newblock \bibinfo{howpublished}{URL:
  \url{https://webscope.sandbox.yahoo.com/catalog.php?datatype=r}}.
\newblock


\bibitem[Yang et~al\mbox{.}(2018)]%
        {yang2018unbiased}
\bibfield{author}{\bibinfo{person}{Longqi Yang}, \bibinfo{person}{Yin Cui},
  \bibinfo{person}{Yuan Xuan}, \bibinfo{person}{Chenyang Wang},
  \bibinfo{person}{Serge Belongie}, {and} \bibinfo{person}{Deborah Estrin}.}
  \bibinfo{year}{2018}\natexlab{}.
\newblock \showarticletitle{Unbiased Offline Recommender Evaluation for
  Missing-Not-At-Random Implicit Feedback}. In
  \bibinfo{booktitle}{\emph{Proceedings of the 12th ACM Conference on
  Recommender Systems}}. \bibinfo{pages}{279--287}.
\newblock


\bibitem[Zhang et~al\mbox{.}(2020)]%
        {zhang2020task}
\bibfield{author}{\bibinfo{person}{Yichi Zhang}, \bibinfo{person}{Zhijian Ou},
  {and} \bibinfo{person}{Zhou Yu}.} \bibinfo{year}{2020}\natexlab{}.
\newblock \showarticletitle{Task-oriented Dialog Systems that Consider Multiple
  Appropriate Responses under the Same Context}. In
  \bibinfo{booktitle}{\emph{Proceedings of the AAAI Conference on Artificial
  Intelligence}}, Vol.~\bibinfo{volume}{34}. \bibinfo{pages}{9604--9611}.
\newblock


\end{thebibliography}

\appendix

\section{Experiments}
\label{exp-setup}

\header{Simulating preference elicitation}
To evaluate the effects of an unbiased recommendation method, ideally, we need an unbiased held-out dataset collected with a randomized logging policy at the item-topic level, free from the effects of selection bias~\citep{huang2020keeping, schnabel2016recommendations}. Unfortunately and to the best of our knowledge, no such dataset exists for \ac{PE}.
As a solution, we propose a simple method to simulate a benchmark dataset to evaluate the effects of selection bias in \ac{PE}. For each topic $t$, we aggregate the ratings from each item $i$ which belongs to the topic, for both the biased training set and the unbiased test set. As a result, we get a biased training set with user-topic interactions and an unbiased test set without the effects to selection bias, to evaluate the performance of various debiasing methods.

\header{Synthetic topic generation}\label{sec:synt-topic}
An item's topic category information is not always guaranteed to be present, for reasons such as privacy constraints from external vendors, noisy or unreliable topic labelling, etc. 
To deal with this issue, we propose a synthetic topic generation method that only relies on user-item interaction information. 
Given user-item interactions, we create a bipartite graph $G= \langle V,E\rangle$, where the set of vertices $V$ is divided into two groups, one of which consists of nodes representing users, and the other has nodes representing items. 
The set $E$ consists of edges between the two groups. 
Each interaction pair $(u,i)$ results in an edge between the node corresponding to $i$ and $u$. 
Given this bipartite-graph, we learn node embeddings via graph representation learning \ac{BINE}~\cite{gao2018bine}.
We make use of a small unbiased test set to generate the bipartite graph, in an attempt to learn unbiased network embeddings. Given the vector representation of all items from the graph embedding method, we use clustering to group the items in the embedding space. 
We use Gaussian mixture models \cite{reynolds2009gaussian} to cluster the embeddings. 
The cluster centers are considered as the topics. 

\header{Coat dataset} This dataset consists of user interactions for a coat-recommendation service, which includes ratings from 290 users on 300 items which are self-selected by users, i.e., these are MNAR ratings~\cite{schnabel2016recommendations}. 
For the unbiased test, a uniformly-random policy is deployed to collect unbiased ratings on 10 items. 
Items are labelled with topics in the dataset, where each item can belong to multiple categories. 
Propensity scores $P(O_{u,i}=1)$ are computed using logistic-regression with item covariates.

\header{Hyperparameters}
We use 5-fold cross-validation for hyper-para\-meter tuning in all our experiments. 
We use Adam \cite{kingma2014adam} for optimizing the model-parameters for the loss-functions defined previously. 
For hyper-parameter tuning, we use the \ac{SNIPS} estimator~\cite{swaminathan2015self}, and optimize for \ac{MAE}. 

\end{document}